**Experimental Measurement of Assembly Indices are Required to Determine The Threshold for Life**

Sara I. Walker,[1]* Cole Mathis,[2] Stuart Marshall,[3] Leroy Cronin[3]*

[1]School of Earth and Space Exploration, Arizona State University, Tempe, Arizona, 85287-0506, USA.

[2]School for Complex Adaptive Systems, Arizona State University, Tempe, Arizona, 85287-0506, USA.

[3]School of Chemistry The University of Glasgow, Glasgow, UK.

Email: sara.i.walker@asu.edu and Lee.Cronin@glasgow.ac.uk

**Abstract**

Assembly Theory (AT) was developed to help distinguish living from non-living systems. The theory is simple as it posits that the amount of selection or Assembly is a function of the number of complex objects where their complexity can be objectively determined using assembly indices. The assembly index of a given object relates to the number of recursive joining operations required to build that object and can be not only rigorously defined mathematically but can be experimentally measured. In pervious work we outlined the theoretical basis, but also extensive experimental measurements that demonstrated the predictive power of AT. These measurements showed that is a threshold in assembly indices for organic molecules whereby abiotic chemical systems could not randomly produce molecules with an assembly index ≥15. In a recent paper by Hazen *et al* [1] the authors not only confused the concept of AT with the algorithms used to calculate assembly indices, but also attempted to falsify AT by calculating theoretical assembly indices for objects made from inorganic building blocks. A fundamental misunderstanding made by the authors is that the threshold is a requirement of the theory, rather than experimental observation. This means that exploration of inorganic assembly indices similarly requires an experimental observation, corelated with the theoretical calculations. Then and only then can the exploration of complex inorganic molecules be done using AT and the threshold for living systems, as expressed with such building blocks, be determined. Since Hazen *et al*.[1] present no experimental measurements of assembly theory, their analysis is not falsifiable.

The recent paper by Hazen *et al.*[1] contains numerous concerning errors and inconsistencies, some of which we outline in this comment. First, the paper title, *'Molecular assembly indices of mineral heteropolyanions: some abiotic molecules are as complex as large biomolecules'* misrepresents the results they present in the manuscript, since the work is hypothetical, presents no empirical data to support the title statement, and furthermore they do not report *any* molecular assembly indices. Instead, the paper does report results of calculations for heteropolyanions that do not support direct comparison to molecular assembly values reported in the literature. No experimental measurements or technical arguments are made to motivate or corroborate the comparison, and indeed no proposal for how to experimentally validate the conclusions of the paper are stated either. As we will discuss below this raises major concerns because the proposed calculations of assembly index for heteropolyanions are not experimentally verifiable and are based on different foundations than molecular assembly.

The abstract of the Hazen paper [1] contains factual errors that mislead the reader. The authors wrote 'A central hypothesis of assembly theory is that any molecule with an assembly index ≥15 found in significant local concentrations represents an unambiguous sign of life.'. This is not correct. The authors make this claim in the abstract (and a similar claim in the introduction) without citation. Indeed, if one accurately reviews the literature, no such hypothesis can be found tying a threshold value for the assembly index uniquely to ~ 15 with any prediction made by assembly theory (AT), including in any of the papers cited by Hazen et al. Instead, what has been reported are experiments confirming the existence of a threshold, where for molecular assembly (e.g., assembly index of covalently bonded molecules) the threshold of MA $\geq$ 15 was empirically identified when no abiotic samples tested exhibited a measured value above this threshold [2-7]. Specifically, the passage in [3] referring to a threshold of 15 states *"These results demonstrate that we can identify the living systems by looking for mixtures with an MA greater than a certain threshold, see Fig. 4.* ***In the case of the analysis presented here it appears that only living samples produced MA measurement above ~15"***. The hypothesis of AT is that a threshold should exist, but the value of MA $\geq$ 15 for covalently bonded molecules was determined by experiment. Therefore, the statement by Hazen et al. regarding the "central hypothesis" of AT is not correct and represents a misreading of the literature and the role of experiment in validating (or invalidating) conjectures of the theory.

Hazen *et al.*[1] do not appropriately cite the peer-reviewed literature. For example, the paper describes an AT inspired investigation of molecule-like heteropolyanions but fails to cite previous work exploring this same idea [8]. Whilst heteropolyanions can be isolated in solution, they are very unstable [9]. In the literature on AT a central tenant repeatedly stated is that to confirm validity for biosignature assessment, the assembly index must be corroborated by its experimental measurement [2-7] (see e.g., [10] for a recent review). Given the instability of heteropolyanions, the assembly indices for inorganic non-covalent systems are likely not experimentally observable. This means that the hypothesis explored by Hazen *et al.*[1] regarding a threshold value of assembly index for life detection in heteropolyanions is itself not testable. A clear example of where this can mislead the reader is in the authors' suggestion their calculations have been applied to "molecules," while no covalently bonded structures are considered in the manuscript. There is an important and physically meaningful difference between measurable molecular structures, and synthetic analogues to hypothesized mineral precursors. The claims of this paper are based on the latter, where the calculations presented include only "molecule-like" heteropolyanions. Comparing the computed values of assembly indices for these heteropolyanions to molecular assembly values of covalently bonded molecules is a category error comparing two things that are only superficially alike. In [3], we have used covalent bonds as building blocks. The literature makes clear how the assembly index can be applied to systems constructed from different sets of building blocks other than covalent bonds, but that one should only compare objects within the same set when discussing how evolution and selection can generate complex objects [7]. The assembly index would be different if we chose atoms, quarks, or some other structural motif to build molecules, and hence a threshold for life detection would be expected to different as explained in ref 7; however, bonds are used because this is what is empirically testable using spectroscopic techniques.[3,5] Therefore, even if the hypothetical assembly indices of the inorganic structures in discussed in Hazen *et al.*[1] were experimentally verifiable, they still could not be directly compared to molecular assembly values for organic molecules because the assembly index values are calculated in a different way, with different building blocks. The hypothesis of assembly theory is that a threshold will exist, but there is no expectation it will be the same for all objects, independent of how they are constructed. We might expect if an assembly theory of minerals can be developed and experimentally verified, that it will have a different threshold for life detection than covalently bonded chemistry.

The paper also states 'AT has not gone unchallenged' citing non-peer reviewed blogs and other commentary which we believe introduces a false narrative. One citation points to a preprint originally posted over 1.5 years ago, which has not been published in a peer-reviewed journal. Another is a blog that is more focused on personal accusations and ad hominin commentary than on scientific content. Furthermore, Hazen et al. draw from the blog a reference to stated similarities between AT and prior efforts, without explanation of those similarities. Again, this draws into question their accurate reading of the literature they cite: they state the results of AT "maybe have been anticipated by other authors" referencing a paper published by the author of the cited blog, which was published one year after the first foundational AT paper [2], and which shares only superficial resemblance to some of the theoretical aspects of AT but none of its empirical predictions or results.

In their conclusion, Hazen *et al.*[1] state 'while the proposal of a biosignature based on a molecular assembly index of 15 is an intriguing and testable concept, the contention that only life can generate molecular structures with MA index ≥ 15 is in error.' The authors concluded their abstract with a similar statement that 'the values of molecular assembly indices >15 do not represent unambiguous biosignatures'. However, this cannot be concluded from their hypothetical calculations, particularly in the absence of any efforts made to connect their conclusion to experimental evidence that could support it. Oddly this statement may also contradict earlier work by the same authors, where minerals claimed to not be biosignatures in the present manuscript form in Stage 7 of 'mineral evolution', which has previously been associated by the very same authors with the era of "bio-mediated mineralogy" [11]. Nonetheless, we contend that claims of whether a given mineral, or molecule, can only form via an evolutionary process (is a sign of life) must be something that can be rigorously tested in the lab, so we regard it as premature to assign "sign of life" to any structure – mineral, molecule or otherwise - in the absence of strong theoretical and empirical grounding of the claim. To date, measurement of organic molecules with MA>15 have only been found to be produced from systems that are associated with life or technology [3-5] and this is motivated by a theoretical explanation [7]. Since the work of Hazen *et al.*[1] is hypothetical, misrepresents the literature, and states conclusions that build on falsities and are not empirically verifiable, we feel that this paper should not have been published. We have written this comment to correct the scientific record and address any confusion.